\PassOptionsToPackage{unicode}{hyperref}
\PassOptionsToPackage{hyphens}{url}
\documentclass[
]{article}
\usepackage{amsmath,amssymb}
\usepackage{lmodern}
\usepackage{iftex}
\ifPDFTeX
  \usepackage[T1]{fontenc}
  \usepackage[utf8]{inputenc}
  \usepackage{textcomp} % provide euro and other symbols
\else % if luatex or xetex
  \usepackage{unicode-math}
  \defaultfontfeatures{Scale=MatchLowercase}
  \defaultfontfeatures[\rmfamily]{Ligatures=TeX,Scale=1}
\fi
\IfFileExists{upquote.sty}{\usepackage{upquote}}{}
\IfFileExists{microtype.sty}{% use microtype if available
  \usepackage[]{microtype}
  \UseMicrotypeSet[protrusion]{basicmath} % disable protrusion for tt fonts
}{}
\makeatletter
\@ifundefined{KOMAClassName}{% if non-KOMA class
  \IfFileExists{parskip.sty}{%
    \usepackage{parskip}
  }{% else
    \setlength{\parindent}{0pt}
    \setlength{\parskip}{6pt plus 2pt minus 1pt}}
}{% if KOMA class
  \KOMAoptions{parskip=half}}
\makeatother
\usepackage{xcolor}
\IfFileExists{xurl.sty}{\usepackage{xurl}}{} % add URL line breaks if available
\IfFileExists{bookmark.sty}{\usepackage{bookmark}}{\usepackage{hyperref}}
\hypersetup{
  hidelinks,
  pdfcreator={LaTeX via pandoc}}
\usepackage{longtable,booktabs,array}
\usepackage{calc} % for calculating minipage widths
\usepackage{etoolbox}
\makeatletter
\patchcmd\longtable{\par}{\if@noskipsec\mbox{}\fi\par}{}{}
\makeatother
\IfFileExists{footnotehyper.sty}{\usepackage{footnotehyper}}{\usepackage{footnote}}
\makesavenoteenv{longtable}
\usepackage{graphicx}
\makeatletter
\def\maxwidth{\ifdim\Gin@nat@width>\linewidth\linewidth\else\Gin@nat@width\fi}
\def\maxheight{\ifdim\Gin@nat@height>\textheight\textheight\else\Gin@nat@height\fi}
\makeatother
\setkeys{Gin}{width=\maxwidth,height=\maxheight,keepaspectratio}
\makeatletter
\def\fps@figure{htbp}
\makeatother
\providecommand{\tightlist}{%
  \setlength{\itemsep}{0pt}\setlength{\parskip}{0pt}}
\ifLuaTeX
  \usepackage{selnolig}  % disable illegal ligatures
\fi
\newlength{\cslhangindent}
\newlength{\csllabelwidth}
\newenvironment{CSLReferences}[2] % #1 hanging-ident, #2 entry spacing
 {% don't indent paragraphs
  \setlength{\parindent}{0pt}
  \ifodd #1 \everypar{\setlength{\hangindent}{\cslhangindent}}\ignorespaces\fi
  \ifnum #2 > 0
  \setlength{\parskip}{#2\baselineskip}
  \fi
 }%
 {}
\usepackage{calc}
\usepackage{tabularx}
\usepackage{authblk}

\title{Point-in-Time Audit Before Alpha: Public-Archive Availability and a Negative Matched-Budget Study on BTC Perpetual Futures}
\author[1]{Baocheng Zeng}
\author[1]{Jinhao Yang}
\author[2]{Peilin Han}
\author[3]{Kangnan He}
\affil[1]{Tsinghua University}
\affil[2]{Beijing University of Posts and Telecommunications}
\affil[3]{Nanjing University}
\date{August 2026}

\begin{document}

\maketitle

\begin{abstract}

Public cryptocurrency archives may appear usable when files exist,
although factor research requires observations available and executable
at each decision time. We audit public Binance BTCUSDT USD-M
perpetual-futures data using event, publication, and availability times
and separate proposal from deterministic auditing, evaluation, and
holdout access. An initial gapless five-minute requirement for trade,
mark, index, and open interest failed: the longest unrepaired
intersection was \textbf{304.5729166666667 days}. A disclosed revision
made trade, mark, index, and realized funding the core streams and made
OI optional because its publication time was unverified. The revised
mask retained \textbf{727 complete UTC days} and supported a mechanical
436/145/146-day train/validation/historical-holdout split.

On 80 frozen known-rule templates, the auditor detected 40/40 violations
and rejected 0/40 legal templates; Wilson 95\% intervals were
91.24\%--100\% and 0\%--8.76\%, so this is template-level evidence.
Across ten null-signal paths, full auditing reduced mean false passes
from 0.2910 to 0.0625 (78.5\%; paired path-bootstrap difference -0.2285,
95\% interval {[}-0.2595, -0.2012{]}). The simulation did not
distinguish exact-grid deletion from availability masking and showed no
material PBO improvement.

Under a shared DSL and matched valid-candidate budgets, a deterministic
adaptive audited agent and random search each produced 39
validation-qualified candidates; tree GP produced 29. In the one-time
historical holdout, all 23 evaluated runs had positive IC but negative
net Sharpe under primary costs, and none of 460 fee--slippage cells had
positive Sharpe. We retain the audit and null-control result but reject
agent superiority, profitability, stability, and an independent masking
effect.

\textbf{Keywords:} perpetual futures; point-in-time data; availability time; factor mining; false discovery; negative result
\end{abstract}

\hypertarget{introduction}{%
\subsection{1. Introduction}\label{introduction}}

Formulaic factor discovery is commonly presented as a search problem.
For perpetual futures, however, a candidate can appear predictive
because the underlying record was unavailable, the trade used an
unfinished bar, realized funding was moved backward, or a gap was filled
with future information. A valid study must therefore establish not only
what an observation represents, but when it could have been used.

We study public BTCUSDT USD-M perpetual archives over
\texttt{{[}2024-08-01,\ 2026-08-01)}. The project began with a stricter
premise: trade, mark, index, and OI had to occupy one exact five-minute
grid continuously for at least 365 days. The checksum-closed archive
audit falsified that premise. The longest unrepaired intersection
contained 87,717 rows, or 304.5729166666667 days. We preserved that
negative finding and revised the admission rule rather than silently
repairing the data. The revised core consists of trade, mark, index, and
realized funding aligned by availability time; OI is optional because
its publication time is unverified.

This study asks:

\begin{enumerate}
\def\labelenumi{\arabic{enumi}.}
\tightlist
\item
  Can the revised core streams support at least 365 complete
  point-in-time days and a frozen temporal split?
\item
  Does a deterministic auditor correctly classify frozen legal and
  known-violation templates?
\item
  Does auditing reduce false passes in null-signal controls?
\item
  Does an adaptive audited proposal policy outperform random search and
  tree GP under identical valid-candidate budgets?
\item
  Do frozen candidates retain predictive and economic value in a
  one-time historical holdout?
\end{enumerate}

Our contributions are deliberately narrow. First, we report both the
failed exact-grid premise and the revised availability-masked admission
result. Second, we integrate three-timestamp rules, executable-plan
auditing, null controls, append-only candidate accounting, and one-time
holdout access. Third, we report the negative search and economic
findings rather than selecting a post-holdout rescue specification. We
do \textbf{not} claim the first safe, auditable, or agentic
factor-mining framework; accepted work already covers formula search,
reinforcement learning, LLM factor agents, and structure-aware
exploration (Ren et al. 2024; Yu et al. 2023; H. Shi et al. 2025; Li et
al. 2024; Chen et al. 2026; Y. Shi, Duan, and Li 2026).

\hypertarget{related-work}{%
\subsection{2. Related Work}\label{related-work}}

RiskMiner uses risk-seeking Monte Carlo tree search for formulaic alpha
discovery (Ren et al. 2024). AlphaGen studies reinforcement learning for
synergistic formula collections (Yu et al. 2023), while AlphaForge
jointly mines and dynamically combines formulaic factors (H. Shi et al.
2025). FAMA evaluates a neural-symbolic LLM factor-mining agent (Li et
al. 2024). AlphaSAGE uses structure-aware GFlowNets and emphasizes
exploration, convergence, runtime, and cross-universe evidence (Chen et
al. 2026). Navigating the Alpha Jungle combines LLM guidance with MCTS
and rich search feedback (Y. Shi, Duan, and Li 2026). These studies
motivate strong baselines, repeated runs, out-of-sample evaluation,
ablation, and economic assessment. Their principal evidence is
equity-centric and does not establish the specific public-archive timing
semantics tested here.

Recent public work also narrows any broad novelty claim. Constrained
cryptocurrency factor agents, safe DSL execution, executable program
evolution, high-frequency digital-asset factor feedback, and
execution-governed perpetual tuning have all been publicly described
(Huang et al. 2026; R. Shi et al. 2026; Lin et al. 2026; Zhang et al.
2026; Deng 2027). Our distinguishing question is consequently not
whether an agent can emit formulas, but whether a perpetual-specific
point-in-time audit changes false passes and whether an adaptive
proposer survives an exactly matched valid-evaluation comparison.

\hypertarget{method}{%
\subsection{3. Method}\label{method}}

\hypertarget{data-source-and-market-level-fallback}{%
\subsubsection{3.1 Data source and market-level
fallback}\label{data-source-and-market-level-fallback}}

Historical OKX probes returned old trade, mark, and index candles but
empty old funding and OI responses through the tested public route. The
frozen fallback rule permitted a whole-study switch, not cross-venue
stream mixing. We therefore used public Binance BTCUSDT USD-M archives
(Binance 2026). The raw manifest contains 826 official-checksum-verified
ZIP files: 24 monthly files for each of trade, mark, index, and funding,
plus 730 daily metrics/OI files.

\hypertarget{event-publication-and-availability-time}{%
\subsubsection{3.2 Event, publication, and availability
time}\label{event-publication-and-availability-time}}

For a bar, event time identifies the interval, publication cannot
precede the explicit close, and availability is
\texttt{close\_time\ +\ 1\ ms}. For realized funding, event time is
settlement time. The archive lacks a distinct publication-time field, so
the primary analysis uses event time plus five minutes as a research
assumption, with 0/5/10/15-minute admission sensitivity. A decision at
time \texttt{t} may read only records with
\texttt{availability\_time\ \textless{}=\ t}.

Missing observations are removed through an explicit mask.
Interpolation, backward filling, cross-gap forward filling, timestamp
snapping, deletion of conflicting duplicates to manufacture continuity,
and venue mixing are prohibited. OI does not determine admission and is
disabled as a factor input because its publication time is unverified.

\hypertarget{protocol-revision-and-split}{%
\subsubsection{3.3 Protocol revision and
split}\label{protocol-revision-and-split}}

The retired rule measured the exact five-minute intersection of trade,
mark, index, and OI. The revised rule measures complete decision days
for trade, mark, index, and native-frequency funding after availability
masking. These are different sample definitions, not two estimates of
one quantity.

The revised data produced 727 complete UTC days. A timestamp-only
60/20/20 split yielded 436 training days, 145 validation days, and 146
test days, with 60-minute purge and embargo. Because the protocol was
hash-locked after the historical interval had ended, the last partition
is called a \textbf{one-time historical holdout}, not a prospective
test.

\hypertarget{deterministic-auditor}{%
\subsubsection{3.4 Deterministic auditor}\label{deterministic-auditor}}

Search methods propose expressions, but a deterministic evaluator
exclusively parses the expression, constructs lineage, applies
availability masks, audits execution, computes labels and metrics,
performs selection, and writes the ledger. The auditor rejects five
known classes:

\begin{enumerate}
\def\labelenumi{\arabic{enumi}.}
\tightlist
\item
  future source shift;
\item
  same-bar execution;
\item
  future realized funding;
\item
  incorrect OI availability alignment; and
\item
  backward fill.
\end{enumerate}

The conformance benchmark contains 40 illegal and 40 legal executable
templates, with eight violations per class. Acceptance required at least
38/40 illegal templates detected, no more than 2/40 legal templates
rejected, and 8/8 exact reason matches in every class. Wilson intervals
accompany, but do not replace, this finite-template rule.

\hypertarget{null-signal-benchmark}{%
\subsubsection{3.5 Null-signal benchmark}\label{null-signal-benchmark}}

The benchmark uses five moving-block label permutations of public
15-minute BTC data and five synthetic heavy-tailed, volatility-clustered
null paths. Each path evaluates 100 candidates, including 80 legal and
20 known-violation templates. We compare no audit, a basic shift audit,
and the full five-class audit under 2\%, 5\%, and 10\% missingness and
three policies: exact-grid deletion, prohibited naive backward fill, and
availability masking. Selection uses BH-FDR \texttt{q=0.10}; PBO and
Deflated Sharpe diagnostics are also recorded. The primary comparison is
5\% missingness with availability masking.

\hypertarget{matched-budget-search}{%
\subsubsection{3.6 Matched-budget search}\label{matched-budget-search}}

All methods share an expression DSL with maximum AST depth 4, at most 5
operator nodes, and lookbacks of 3/6/12/24/48/96/288 five-minute bars. A
numerically valid candidate must pass the audit, be unique, have at
least 0.95 finite coverage, and have finite train and validation IC.
Later qualification requires BH-FDR \texttt{q\textless{}0.10}, a
dependence-aware interval excluding zero, and performance above the
random-search 90th percentile, followed by
\texttt{\textbar{}correlation\textbar{}\textless{}0.8} deduplication.

We compare:

\begin{itemize}
\tightlist
\item
  \textbf{audited agent:} a deterministic adaptive proposal policy using
  training-only failure and validation feedback; it is not an external
  LLM;
\item
  \textbf{random:} uniform proposals in the same DSL;
\item
  \textbf{tree GP:} subtree mutation and crossover under the same DSL
  and validity gate.
\end{itemize}

The seeds are 11, 23, 37, 53, and 71; horizons are 15 and 60 minutes.
Each method--seed--horizon run evaluates 100 valid candidates, giving
1,000 valid candidates per method. All successful, invalid, duplicate,
and rejected proposals are retained.

\hypertarget{frozen-holdout-and-economic-evaluation}{%
\subsubsection{3.7 Frozen holdout and economic
evaluation}\label{frozen-holdout-and-economic-evaluation}}

At most five qualified factors per run are direction-corrected on
validation data, z-scored, and equally weighted. Position is the sign of
the composite, delayed by one completed decision bar and sampled
non-overlapping at the target horizon. The primary cost model uses a
conservative Binance research assumption of 6 bp taker fee per side plus
2 bp slippage per side. Fee sensitivity is 4/5/6/8/10 bp and slippage
sensitivity is 1/2/5/10 bp per side. Realized funding is charged only
when a position crosses a settlement event.

The holdout evaluator was invoked once after selection and economic
rules were frozen. Candidate-level IC intervals use 500 moving-block
resamples. Post-holdout method summaries use a descriptive
10,000-resample seed-cluster bootstrap that retains both horizons for a
sampled seed.

\hypertarget{experiments}{%
\subsection{4. Experiments}\label{experiments}}

We organize the experiments by the five research questions.

\begin{itemize}
\tightlist
\item
  \textbf{Data audit:} census coverage, duplicates, ordering defects,
  exact-grid gaps, funding cadence, and lag sensitivity.
\item
  \textbf{Template audit:} confusion matrix and reason-code accuracy on
  80 frozen programs.
\item
  \textbf{False-discovery controls:} 10 primary null paths and 270
  audit-by-gap scenarios.
\item
  \textbf{Search comparison:} 3 methods $\times$ 5 seeds $\times$ 2 horizons $\times$ 100
  valid candidates.
\item
  \textbf{Historical holdout:} frozen candidates, primary costs, 20
  fee--slippage combinations per evaluated run, one-bar delay,
  extreme-observation deletion, and volatility-state checks.
\end{itemize}

The principal outcomes are complete days; violation recall and legal
false rejection; false-pass rate and PBO; valid and qualified
candidates; test IC and RankIC; net Sharpe, cumulative return, drawdown,
turnover, and robustness counts. Direction accuracy, trade count, and
evaluations to first discovery were not emitted upstream and are not
reconstructed.

% Compact and balance float-only pages: fixed inter-float spacing with
% symmetric stretch above and below the complete float group.
\makeatletter
\setlength{\@fptop}{0pt plus 1fil}
\setlength{\@fpsep}{12pt plus 2pt minus 2pt}
\setlength{\@fpbot}{0pt plus 1fil}
\makeatother

\hypertarget{results}{%
\subsection{5. Results}\label{results}}

\hypertarget{public-archive-availability}{%
\subsubsection{5.1 Public-archive
availability}\label{public-archive-availability}}

\begin{longtable}[]{@{}llrrl@{}}
\caption{Native-stream archive coverage and point-in-time admission
roles.}\tabularnewline
\toprule
Stream & Admission role & Rows/events & Dates & Availability rule \\
\midrule
\endfirsthead
\toprule
Stream & Admission role & Rows/events & Dates & Availability rule \\
\midrule
\endhead
Trade & Core & 210,240 & 730 & close + 1 ms \\
Mark & Core & 209,952 & 729 & close + 1 ms \\
Index & Core & 209,952 & 729 & close + 1 ms \\
Funding & Core & 2,190 & 730 & event + 5 min assumption \\
OI & Optional; disabled & 210,235 & 730 & unverified \\
\bottomrule
\end{longtable}

Trade had no exact-grid gap. Mark and index each lacked 288 bars on
2026-06-29. Funding had no separate publication-time field. OI had 9
missing and 3 off-grid records, one conflicting duplicate group, and 501
archive-order reversals; these defects were retained rather than
repaired.

The old trade/mark/index/OI exact-grid requirement failed at
304.5729166666667 continuous days. Under the revised core definition,
the five-minute primary mask retained 209,951 eligible decisions, masked
288, and yielded 727 complete UTC days. Funding lag 0/5/10/15 minutes
left the complete-day count at 727; longest eligible runs were
697.000/697.000/696.997/696.993 days. This supports data admission only,
not alpha.

\begin{figure}
\centering
\includegraphics[width=0.85\textwidth,height=\textheight]{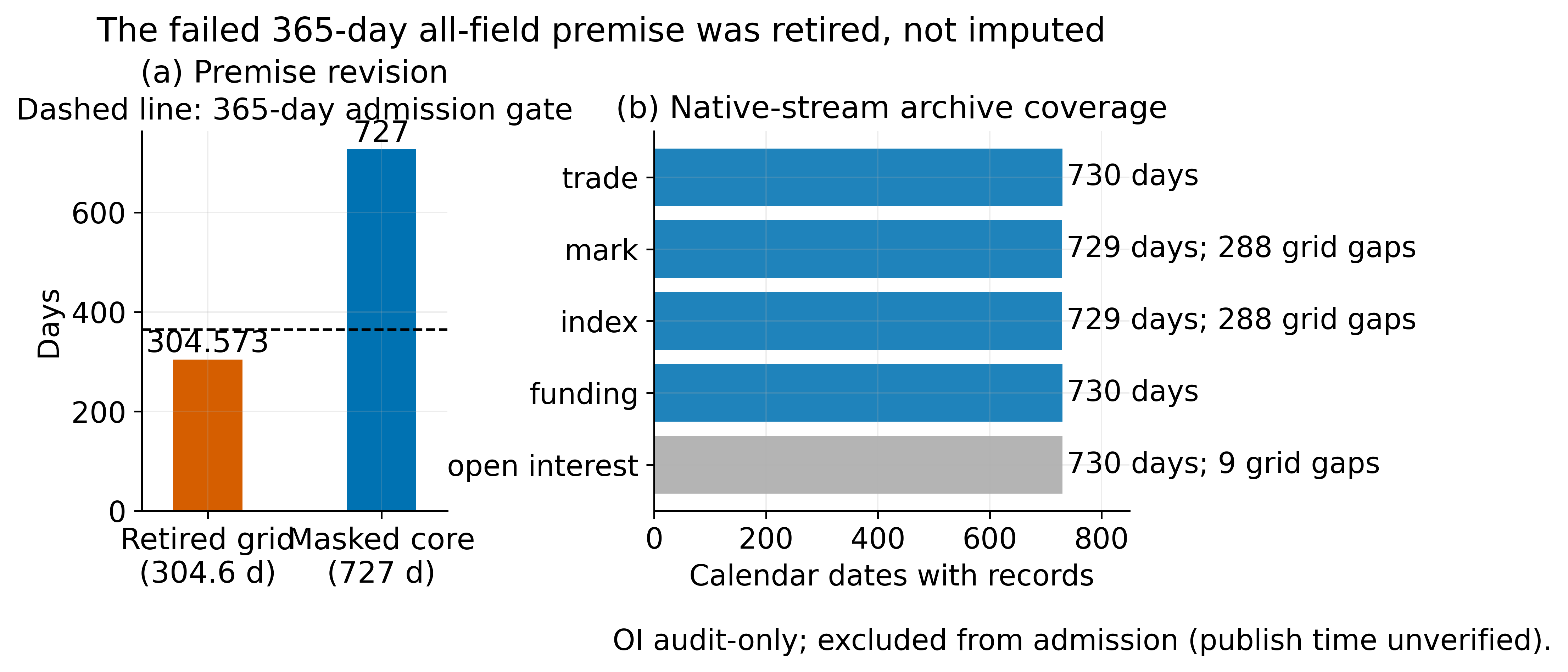}
\caption{Retired exact-grid and revised masked-core availability
(different sample definitions); the dashed line marks the 365-day
admission gate}
\end{figure}

\hypertarget{known-template-audit}{%
\subsubsection{5.2 Known-template audit}\label{known-template-audit}}

\begin{table}[htbp]
\centering
\caption{Known-template auditor conformance with Wilson uncertainty.}
\footnotesize
\setlength{\tabcolsep}{3.5pt}
\begin{tabularx}{\linewidth}{@{}>{\raggedright\arraybackslash}Xrrr@{}}
\hline
Scope & \shortstack{Detected or\\rejected} & Rate & \shortstack{Wilson 95\%\\interval} \\
\hline
Illegal templates & 40/40 & 1.000 & [0.9124, 1.0000] \\
Legal templates falsely rejected & 0/40 & 0.000 & [0.0000, 0.0876] \\
Each violation class & 8/8 & 1.000 & [0.6756, 1.0000] \\
\hline
\end{tabularx}
\end{table}

All 40 rejection reasons matched the injected class. The point estimates
pass the frozen template rule, but the intervals do not establish
population recall above 95\% or false rejection below 5\%. The generator
directly exercises known auditor rules, so unknown leakage is outside
the evidence.

\hypertarget{null-controls-and-missing-policy-ablation}{%
\subsubsection{5.3 Null controls and missing-policy
ablation}\label{null-controls-and-missing-policy-ablation}}

At 5\% missingness with availability masking, mean false passes were
0.2910 without auditing, 0.2479 with the basic audit, and 0.0625 with
the full audit. Full versus no audit improved all 10 paths; the paired
difference was -0.2285 with a path-bootstrap 95\% interval {[}-0.2595,
-0.2012{]} and exact two-sided sign-test \texttt{p=0.001953}. The
relative reduction was 78.5\%.

Conditional holdout-failure differences were uncertain because full
auditing often selected zero or few candidates. Failure burden per
evaluated candidate fell by 0.1062, interval {[}-0.1237, -0.0865{]}. PBO
changed by only -0.001429, interval {[}-0.004286, 0{]}, with nine ties.
DSR was zero where computable and missing when no candidate was
selected.

\begin{longtable}[]{@{}lrrr@{}}
\caption{Mean false-pass rates at 5\% missingness by audit and
missing-data policy.}\tabularnewline
\toprule
Audit & Exact-grid drop & Backward fill & Availability mask \\
\midrule
\endfirsthead
\toprule
Audit & Exact-grid drop & Backward fill & Availability mask \\
\midrule
\endhead
None & 0.2910 & 0.2890 & 0.2910 \\
Basic & 0.2479 & 0.2521 & 0.2479 \\
Full & 0.0625 & 0.0525 & 0.0625 \\
\bottomrule
\end{longtable}

Exact-grid deletion and availability masking were identical in the
executed simulation, while backward fill was not uniformly worse.
Therefore the supported result is the joint effect of the full
known-rule audit pipeline, not an independent effect of masking.

% Keep the paired Results figures on ordinary text pages rather than a float-only page.
\makeatletter\def\fps@figure{!ht}\makeatother

\begin{figure}
\centering
\includegraphics[width=0.82\textwidth,height=\textheight]{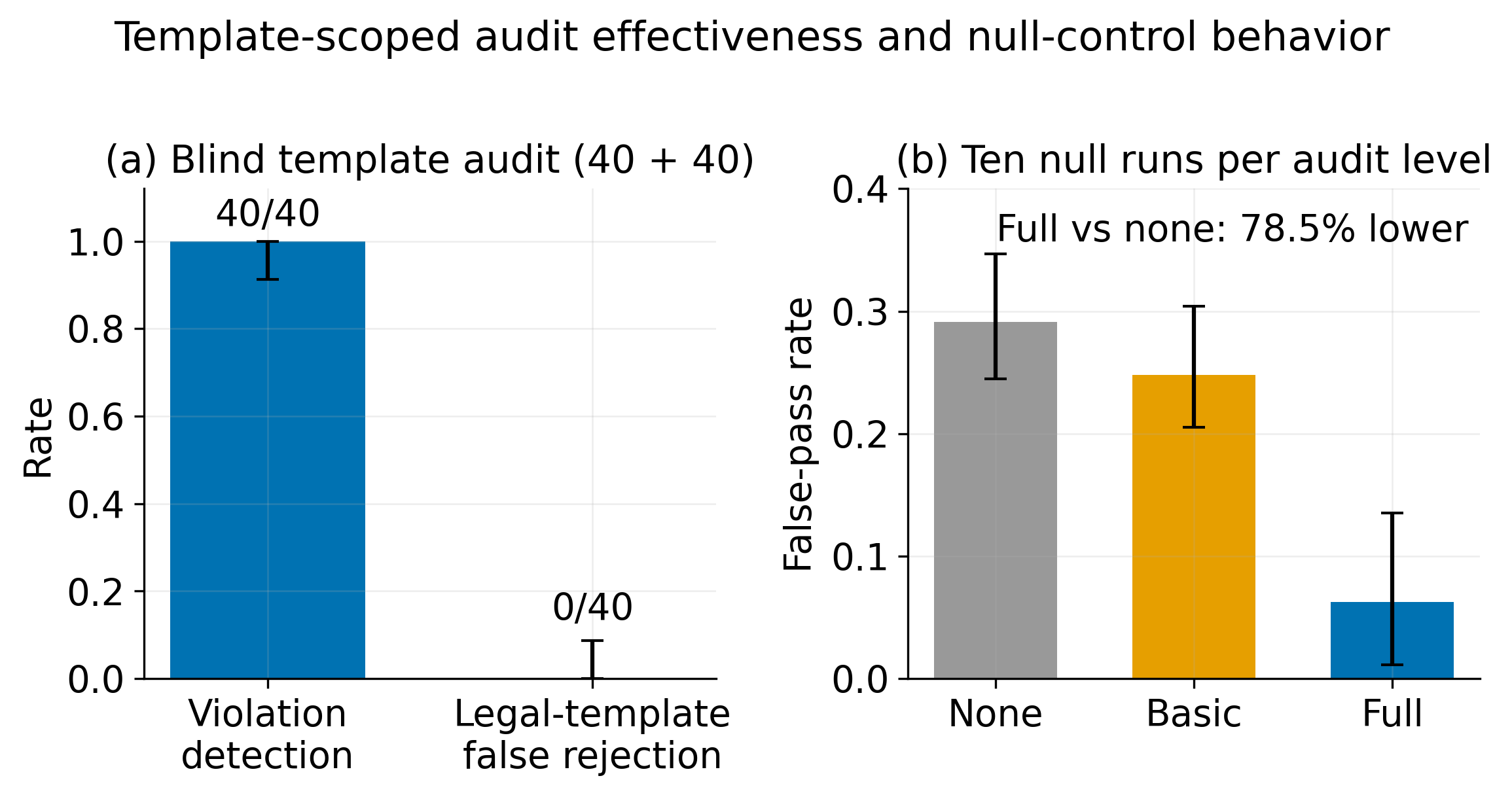}
\caption{Template audit and null controls}
\end{figure}

\hypertarget{search-accounting-and-efficiency-adjudication}{%
\subsubsection{5.4 Search accounting and efficiency
adjudication}\label{search-accounting-and-efficiency-adjudication}}

\begin{table}[htbp]
\centering
\caption{Matched-budget search accounting and qualification yield.}
\footnotesize
\setlength{\tabcolsep}{3.5pt}
\begin{tabularx}{\linewidth}{@{}>{\raggedright\arraybackslash}Xrrrrr>{\raggedleft\arraybackslash}p{76pt}@{}}
\hline
Method & Runs & Attempts & Valid & Qualified & \shortstack{Runs\\selected} & \shortstack{Yield/100 valid\\(95\% CI)} \\
\hline
Audited agent & 10 & 1,648 & 1,000 & 39 & 8 & 3.9 [1.9, 5.0] \\
Random & 10 & 2,658 & 1,000 & 39 & 8 & 3.9 [1.9, 5.0] \\
Tree GP & 10 & 1,804 & 1,000 & 29 & 8 & 2.9 [1.1, 4.7] \\
\hline
\end{tabularx}
\end{table}

The audited agent tied random search on qualified candidates and
therefore did not beat both baselines. Raw attempts were lower for the
agent, but evaluations to first qualified factor and resource-normalized
runtime were not frozen outcomes. We reject search-superiority and
general efficiency claims.

\begin{figure}
\centering
\includegraphics[width=0.82\textwidth,height=\textheight]{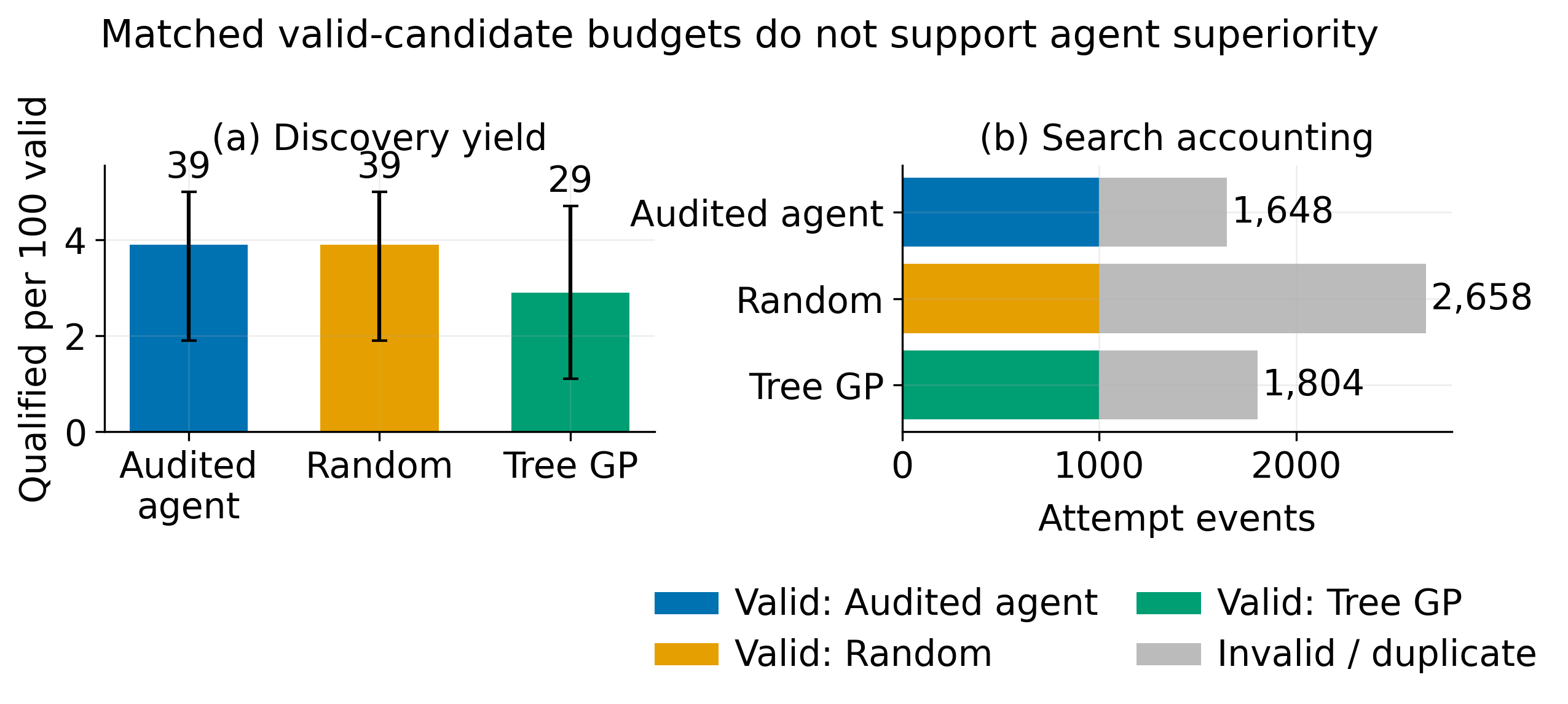}
\caption{Matched-budget search accounting: blue, orange, and green
denote matched-valid audited-agent, random, and tree-GP candidates; gray
denotes invalid or duplicate attempts}
\end{figure}

% Restore the journal template's normal figure placement policy.
\makeatletter\def\fps@figure{htbp}\makeatother

\hypertarget{historical-holdout-prediction-and-economics}{%
\subsubsection{5.5 Historical-holdout prediction and
economics}\label{historical-holdout-prediction-and-economics}}

\begin{table}[htbp]
\centering
\caption{Historical-holdout predictive and economic results at primary costs.}
\small
\setlength{\tabcolsep}{6pt}
\textit{(a) Predictive results}\\[2pt]
\begin{tabular}{lrrr}
\hline
Method & Runs (eval./none) & Mean test IC (95\% CI) & Mean RankIC \\
\hline
Audited agent & 8/2 & 0.2352 [0.1551, 0.3552] & 0.2136 \\
Random & 8/2 & 0.1907 [0.1164, 0.2681] & 0.1826 \\
Tree GP & 7/3 & 0.1550 [0.1262, 0.1765] & 0.1375 \\
\hline
\end{tabular}

\vspace{5pt}
\textit{(b) Economic results}\\[2pt]
\begin{tabular}{lrrrr}
\hline
Method & Positive Sharpe & Median Sharpe & Mean DD & Turnover \\
\hline
Audited agent & 0/8 & -28.04 & -0.9291 & 6,235.9 \\
Random & 0/8 & -21.16 & -0.9460 & 6,652.8 \\
Tree GP & 0/7 & -30.11 & -0.9008 & 5,158.1 \\
\hline
\end{tabular}
\end{table}

Validation-to-test IC sign was retained in all 23 evaluated runs, and
individual moving-block IC intervals excluded zero. Nevertheless, no
evaluated run had positive net Sharpe at primary costs. An additional
one-bar delay produced 0/23 positive Sharpe, and no run remained
economically positive after the frozen extreme-observation deletions.

The fee--slippage grid contained 160 agent, 160 random, and 140 GP
cells. Positive-Sharpe counts were 0, 0, and 0. The best observed Sharpe
values were -6.3148, -5.5009, and -5.4938, respectively. Thus all 460
executed cost cells were negative. Positive IC did not imply an
economically usable strategy.

% Keep the final Results table and figure with adjacent Results/Discussion text.
\makeatletter\def\fps@figure{!ht}\makeatother

\begin{figure}
\centering
\includegraphics[width=0.82\textwidth,height=\textheight]{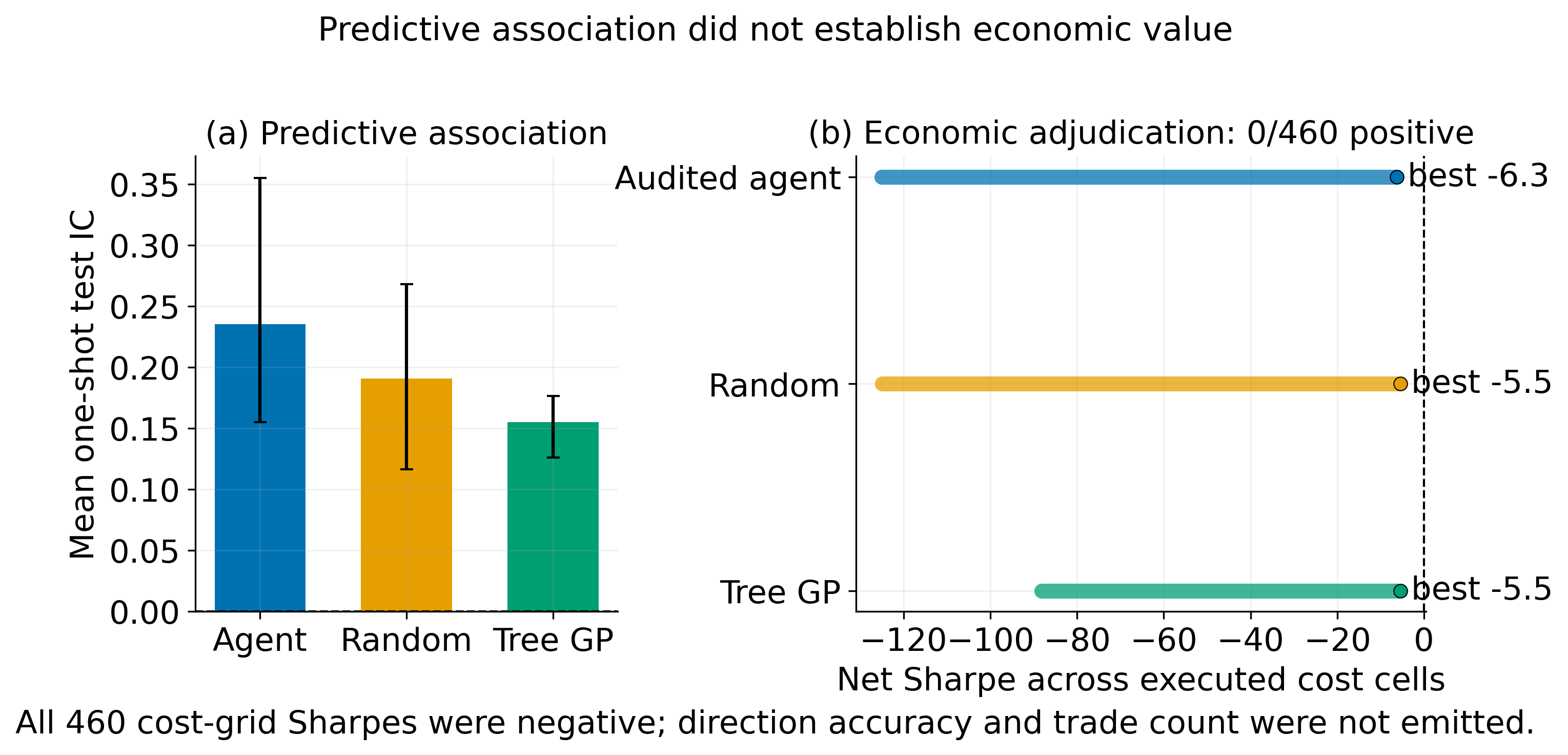}
\caption{Predictive and economic adjudication}
\end{figure}

\makeatletter\def\fps@figure{htbp}\makeatother

\hypertarget{discussion}{%
\subsection{6. Discussion}\label{discussion}}

The strongest positive result is procedural: known-rule auditing removed
candidates that should not enter statistical selection and sharply
reduced false passes in the executed null controls. The result is
relevant because archive completeness alone concealed both an invalid
original data premise and concrete OI and bar gaps. Yet the benchmark is
not an unseen-adversary test, and its finite-template intervals are
wide.

The experiment also separates prediction from economics. Positive
holdout IC coexisted with very high turnover and uniformly negative
cost-adjusted results. Changing the combiner, holding period, or cost
model after opening the holdout could have produced a more attractive
retrospective specification, but would invalidate the one-time-access
design. We retained the negative result.

The adaptive proposer's tie with random search is likewise informative.
Its lower raw-attempt count does not establish the preregistered
discovery-efficiency claim, especially because first-discovery and
resource measurements were not emitted. The result applies only to this
deterministic adaptive policy, DSL, data, and budget; it is not evidence
that all LLM or agentic search is ineffective.

\hypertarget{limitations}{%
\subsection{7. Limitations}\label{limitations}}

\begin{enumerate}
\def\labelenumi{\arabic{enumi}.}
\tightlist
\item
  \textbf{Single asset and venue.} Evidence is limited to Binance
  BTCUSDT USD-M and cannot establish cross-market validity.
\item
  \textbf{Retrospective lock.} The holdout interval ended before the
  protocol lock; one-time access is not prospective validation.
\item
  \textbf{Funding-time assumption.} The archive has no separate
  publication timestamp. The 5-minute assumption was tested for
  admission, not predictive outcomes.
\item
  \textbf{OI excluded.} OI publication time is unverified and its
  archive contains gaps, off-grid records, a conflict, and ordering
  reversals.
\item
  \textbf{Template scope.} The audit benchmark covers known injected
  rules, not unknown or adversarial leakage implementations.
\item
  \textbf{Gap-mechanism identification.} The null simulation did not
  distinguish exact-grid deletion from availability masking.
\item
  \textbf{Incomplete search curves.} Budgets 30 and 300, first-discovery
  counts, and a resource-normalized comparison were not executed.
\item
  \textbf{Missing baseline.} LightGBM was unavailable; the broader
  planned nonlinear and manual baselines were not completed.
\item
  \textbf{Execution approximation.} Aggregated bars and assumed slippage
  do not support capacity or live-tradability claims. Direction accuracy
  and trade count were not emitted.
\item
  \textbf{No external validation.} There is no second asset, exchange,
  or genuinely future interval.
\end{enumerate}

\hypertarget{conclusion}{%
\subsection{8. Conclusion}\label{conclusion}}

A file-complete public archive was not automatically a
point-in-time-complete research dataset. The original exact-grid
requirement failed at 304.5729166666667 days; a disclosed,
scientifically different core-stream definition yielded 727 complete
days without filling. The deterministic auditor passed a finite
known-template benchmark and reduced false passes across the executed
null controls, but the missing-policy ablation did not identify an
independent masking effect. Under matched valid-candidate budgets, the
audited adaptive proposer tied random search and did not establish
superiority. Positive historical-holdout IC failed every primary and
sensitivity economic test.

The evidence supports an audit-focused negative-result paper, not a
profitability or general agent-superiority claim. Detailed protocols,
additional tables, artifact hashes, and exact reproduction commands are
provided in the supplement and reproducibility record.

\small
\interlinepenalty=10000
\clubpenalty=10000
\widowpenalty=10000

\hypertarget{refs}{}
\begin{CSLReferences}{1}{0}
\leavevmode\vadjust pre{\hypertarget{ref-binancepublicdata}{}}%
Binance. 2026. {``Binance Public Data.''} Public archive.
\url{https://data.binance.vision/}.

\leavevmode\vadjust pre{\hypertarget{ref-alphasage2026}{}}%
Chen, Binqi, Hongjun Ding, Ning Shen, Jinsheng Huang, Taian Guo, Luchen
Liu, and Ming Zhang. 2026. {``AlphaSAGE: Structure-Aware Alpha Mining
via GFlowNets for Robust Exploration.''} In \emph{The Fourteenth
International Conference on Learning Representations}.
\url{https://openreview.net/forum?id=zRKF4ln2VE}.

\leavevmode\vadjust pre{\hypertarget{ref-autoquant2026}{}}%
Deng, Kaihong. 2027. {``AutoQuant: An Auditable Expert-System Framework
for Execution-Constrained Auto-Tuning in Cryptocurrency Perpetual
Futures.''} \emph{Expert Systems with Applications} 333: 133924.
\url{https://doi.org/10.1016/j.eswa.2026.133924}.

\leavevmode\vadjust pre{\hypertarget{ref-hypotheses2026}{}}%
Huang, Yikuan, Zheqi Fan, Kaiqi Hu, and Yifan Ye. 2026. {``From
Hypotheses to Factors: Constrained LLM Agents in Cryptocurrency
Markets.''} \url{https://arxiv.org/abs/2604.26747}.

\leavevmode\vadjust pre{\hypertarget{ref-fama2024}{}}%
Li, Zhiwei, Ran Song, Caihong Sun, Wei Xu, Zhengtao Yu, and Ji-Rong Wen.
2024. {``Can Large Language Models Mine Interpretable Financial Factors
More Effectively? A Neural-Symbolic Factor Mining Agent Model.''} In
\emph{Findings of the Association for Computational Linguistics: ACL
2024}, 3891--3902. Bangkok, Thailand: Association for Computational
Linguistics. \url{https://doi.org/10.18653/v1/2024.findings-acl.233}.

\leavevmode\vadjust pre{\hypertarget{ref-factorengine2026}{}}%
Lin, Qinhong, Ruitao Feng, Yinglun Feng, Zhenxin Huang, Yukun Chen,
Zhongliang Yang, Linna Zhou, Binjie Fei, Jiaqi Liu, and Yu Li. 2026.
{``FactorEngine: A Program-Level Knowledge-Infused Factor Mining
Framework for Quantitative Investment.''}
\url{https://arxiv.org/abs/2603.16365}.

\leavevmode\vadjust pre{\hypertarget{ref-riskminer2024}{}}%
Ren, Tao, Ruihan Zhou, Jinyang Jiang, Jiafeng Liang, Qinghao Wang, and
Yijie Peng. 2024. {``RiskMiner: Discovering Formulaic Alphas via Risk
Seeking Monte Carlo Tree Search.''} In \emph{Proceedings of the 5th ACM
International Conference on AI in Finance}, 752--60. Association for
Computing Machinery. \url{https://doi.org/10.1145/3677052.3698613}.

\leavevmode\vadjust pre{\hypertarget{ref-alphaforge2025}{}}%
Shi, Hao, Weili Song, Xinting Zhang, Jiahe Shi, Cuicui Luo, Xiang Ao,
Hamid Arian, and Luis Angel Seco. 2025. {``AlphaForge: A Framework to
Mine and Dynamically Combine Formulaic Alpha Factors.''}
\emph{Proceedings of the AAAI Conference on Artificial Intelligence} 39
(12): 12524--32. \url{https://doi.org/10.1609/aaai.v39i12.33365}.

\leavevmode\vadjust pre{\hypertarget{ref-hubble2026}{}}%
Shi, Runze, Shengyu Yan, Yuecheng Cai, and Chengxi Lv. 2026. {``Hubble:
An LLM-Driven Agentic Framework for Safe, Diverse, and Reproducible
Alpha Factor Discovery.''} \url{https://arxiv.org/abs/2604.09601}.

\leavevmode\vadjust pre{\hypertarget{ref-alphajungle2026}{}}%
Shi, Yu, Yitong Duan, and Jian Li. 2026. {``Navigating the Alpha Jungle:
An LLM-Powered MCTS Framework for Formulaic Alpha Factor Mining.''}
\emph{Proceedings of the AAAI Conference on Artificial Intelligence} 40
(2): 997--1005. \url{https://doi.org/10.1609/aaai.v40i2.37069}.

\leavevmode\vadjust pre{\hypertarget{ref-alphagen2023}{}}%
Yu, Shuo, Hongyan Xue, Xiang Ao, Feiyang Pan, Jia He, Dandan Tu, and
Qing He. 2023. {``Generating Synergistic Formulaic Alpha Collections via
Reinforcement Learning.''} In \emph{Proceedings of the 29th ACM SIGKDD
Conference on Knowledge Discovery and Data Mining}, 5476--86.
Association for Computing Machinery.
\url{https://doi.org/10.1145/3580305.3599831}.

\leavevmode\vadjust pre{\hypertarget{ref-quantevolver2026}{}}%
Zhang, Lingzhe, Tong Jia, Yunpeng Zhai, Zixuan Xie, Chiming Duan,
Minghua He, Philip S. Yu, and Ying Li. 2026. {``From Feedback Loops to
Policy Updates: Reinforcement Fine-Tuning for LLM-Based Alpha Factor
Discovery.''} \url{https://arxiv.org/abs/2605.15412}.

\end{CSLReferences}

\end{document}